\documentclass[a4paper,fleqn,usenatbib]{mnras}

\usepackage{newtxmath}

\usepackage{graphicx}	
\usepackage{amsmath}	
\usepackage{amssymb}	
\usepackage{amsfonts}
\usepackage{multirow}
\usepackage{threeparttablex}
\usepackage{xcolor}
\colorlet{RED}{red}    

\title[A new bright VHS z=6.82 quasar]{A new bright z=6.82 quasar discovered with VISTA: \\ VHS J0411-0907}

\author[E. Pons et al.]
{\parbox{\textwidth}
{E. Pons$^{1,2}$\thanks{E-mail: epons@ast.cam.ac.uk},
R. G. McMahon$^{1,2}$,
R. A. Simcoe$^{3}$,
M. Banerji$^{1,2}$,
P. C. Hewett$^{1}$
and S. L. Reed$^{4}$
}
\\ \\
$^{1}$Institute of Astronomy, University of Cambridge, Madingley Road, Cambridge CB3 0HA, UK\\
$^{2}$Kavli Institute for Cosmology, University of Cambridge, Madingley Road, Cambridge CB3 0HA, UK\\
$^{3}$MIT-Kavli Institute for Astrophysics and Space Research, 77 Massachusetts Ave., Cambridge, MA 02139, US\\
$^{4}$Department of Astrophysical Sciences, Princeton University, 4 Ivy Lane, Princeton, NJ 08544, US
}

\date{Accepted XXX. Received YYY; in original form ZZZ}

\pubyear{2018}

\begin{document}
\label{firstpage}
\pagerange{\pageref{firstpage}--\pageref{lastpage}}
\maketitle


\begin{abstract}
We present the discovery of a new $z \sim 6.8$ quasar discovered with the near-IR VISTA Hemisphere Survey (VHS) which has been spectroscopically confirmed by the ESO New Technology Telescope (NTT) and the Magellan telescope. This quasar has been selected by spectral energy distribution (SED) classification using near infrared data from VISTA, optical data from Pan-STARRS, and mid-IR data from WISE. The SED classification algorithm is used to statistically rank two classes; foreground Galactic low-mass stars and high redshift quasars, prior to spectroscopic observation. Forced photometry on Pan-STARRS pixels for VHS J0411-0907 allows to improve the SED classification reduced-$\chi^2$ and photometric redshift. VHS J0411-0907 ($z=6.82$, $y_{AB} = 20.1$ mag, $J_{AB} = 20.0$ mag) has the brightest J-band continuum magnitude of the nine known quasars at $z > 6.7$ and is currently the highest redshift quasar detected in the Pan-STARRS survey. This quasar has one of the lowest black hole mass ($M_{\rm{BH}}= (6.13 \pm 0.51)\times 10^8\:\mathrm{M_{\odot}}$) and the highest Eddington ratio ($2.37\pm0.22$) of the known quasars at $z>6.5$. The high Eddington ratio indicates that some very high-$z$ quasars are undergoing super Eddington accretion. We also present coefficients of the best polynomials fits for colours vs spectral type on the Pan-STARRS, VISTA and WISE system for MLT dwarfs and present a forecast for the expected numbers of quasars at $z>6.5$.
\end{abstract}




\section{Introduction}

Quasars are the most luminous non-transient sources in the sky making them detectable even in the first billion years of the Universe. They provide important information on the formation and evolution of supermassive black holes (SMBH) and their host galaxies. Also via spectroscopic studies of intervening absorption lines from neutral hydrogen, high-redshift quasars are one of the best objects to study the early universe as tracers of the {\it `Epoch of Reionization'} (EoR). 

Recent studies of the Cosmic Microwave Background (CMB) by the \citet{Planck16} set the beginning of reionization at redshift $z\sim 9$.
On the other hand, quasar spectra can be used to study the intergalactic medium (IGM) at the early stage of the Universe by giving an indication of the fraction of neutral hydrogen of the IGM through the Lyman alpha forest. The study of high-redshift quasars suggests that the reionization was complete by $z\sim6$ \citep{Fan06, Becker07}. In order to get a better constraint on the end the transition phase from neutral to ionized hydrogen, a larger sample of quasars at $z>6.5$ is needed.

The last few years have seen immense progress in the discovery of $z>6.0-6.5$ quasars mainly selected based on optical colours. The search of $z\ga 6.5$ is much more challenging with only a few objects known, and has been successful due to multi-wavelenght observations combining optical, near-infrared (NIR) and mid-infrared (MIR) broadband photometry. In the rest frame spectra of high-redshift quasars neutral hydrogen atoms within the highly ionized IGM absorb all the flux blueward the Ly$\alpha$ emission-line (at rest-frame wavelength of $\lambda_{rest} =1215.67\mathrm{\AA}$) leading to large break between i and z-bands for $z>6$ quasars (i-dropout) and between z and y-bands for $z>6.4$ quasars (z-dropout). However, i-dropout or z-dropout selection also selects a significant foreground galactic population of stellar contaminants which are ultra-cool stars (of spectral type , M, L or T) that can be identified based on their infrared (IR) colours.

Currently $\sim$100 $z>6.0$ quasars have been found mainly thanks to large-area optical and IR surveys such as the Sloan Digital Sky Survey \citep{Fan01, Fan03, fan04, Fan06, Jiang08, Jiang09, Jiang15, Jiang16, Wang16}, the Canada-France-Hawaii Telescope High-z Quasars Surveys \citep[CFHQS;][]{Willott07, Willott09, Willott10}, the UK Infrared Deep Sky Survey \citep[UKIDSS, including the two highest redshift quasars known;][]{Mortlock09, Banados18}, the Pan-STARRS1 distant quasars survey \citep{Banados14, Banados16, Venemans15a, Tang17, Mazzucchelli17}, the VST ATLAS survey \citep{Carnall15, Chehade18} and the the Subaru High-z Exploration of Low-Luminosity Quasars survey \citep[SHELLQS;][]{Matsuoka16}. In addition, deep surveys such as the Dark Energy Survey \citep[DES;]{Reed15, Reed17, Reed19}, the DECam Legacy Survey \citep[DECaLS;][]{Wang17} and the VISTA Kilo-Degree Infrared Galaxy (VIKING) Survey \citep{Venemans13} have been successful in the search of high-redshift quasars, in particular at redshift $z \ga 6.5$.

The presence of supermassive black holes with masses of $\sim 10^9\;\mathrm{M_{\odot}}$ poses a challenge for models of black-hole formation and growth and can potentially place interesting constraints on the seed masses from which these black-holes are assembled. While many models have been proposed for black-hole formation (see e.g. \citet{Volonteri10} for a review), these models often postulate different seed masses from which $z>6$ quasars may be assembled. For example, growing $10^9 - 10^{10}\;\mathrm{M_{\odot}}$ black holes at $z>6$ from Population II/III stellar mass seeds ($\sim 100\;\mathrm{M_{\odot}}$), requires almost continuous growth at the Eddington limit, or alternatively episodes of super-Eddington growth \citep{Alexander14, Volonteri15, Lupi16, Inayoshi16, Trakhtenbrot17}. On the other hand, formation pathways originating in direct collapse of gas clouds require very massive seeds of $\sim 10^5-10^6\; \mathrm{M_{\odot}}$ to be formed at very early times \citep{Begelman06, Inayoshi18, Mayer19}.

In this paper we extend a SED classification method presented in \citet{Reed17} to the Pan-STARRS1 survey to select and classify high-z quasar candidates using optical, NIR, and MIR photometry. We report the discovery of the spectroscopically confirmed highest redshift Pan-STARRS (PS1) quasar, VHS J0411-0907\footnote{an independent discovery using DECaLS and PS1 has been recently reported by \citet{Wang18}}, and the brightest object in the NIR $J$-band among the 9 known quasars at redshift $z>6.7$. The paper is organised as follows. The data used in this work are described in section \ref{sec:data}. The quasar selection method is explained in section \ref{sec:QSOsel} and the discovery of the new high-z quasar is presented in section \ref{sec:vhs0411}. Finally in section \ref{sec:BDmagellan} we report the discovery of two new brown dwarfs which had a high probability of being a quasar based on our classification, demonstrating the current limitation of our method, and in section \ref{sec:numQSO} we give a prediction on the number of high-$z$ quasars that should be in the VHS-PS1 region.

The emission-line wavelengths given in this paper correspond to vacuum wavelengths and all the magnitudes reported are in the AB system. The conversion from Vega to AB that we used for VISTA and WISE are $J_{AB}=J_{Vega}+0.937$, $H_{AB}=H_{Vega}+1.384$, $Ks_{AB}=Ks_{Vega}+1.839$, $W1_{AB}=W1_{Vega}+2.699$ and $W2_{AB}=W2_{Vega}+3.339$. We used the $\rm \Lambda$CDM cosmology with $\Omega_{m0}=0.3$ and $H_0=70.0\;\mathrm{km/s/Mpc}$.

\section{Photometric data}
\label{sec:data}

We used three photometric imaging survey datasets for this work:
Pan-STARRS1 data for the optical, VISTA Hemisphere Survey (VISTA-VHS)
for the NIR and the \textit{Wide-field Infrared Survey Explorer} (WISE)
for the MIR.

\subsection{The VISTA Hemisphere Survey}

The near IR VISTA surveys use the 4m VISTA telescope \citep{Sutherland15} situated at ESO's Paranal Observatory in Chile to conduct NIR imaging of different areas of the sky at different depths. It is using the VIRCAM camera \citep{Dalton06, Gonzalez18} which consists of 16 2k x 2k MCT detectors (chips) in a sparse filled mosaic covering 0.59 deg$^2$ and which is taking 16-non continuous image of the sky, forming a \textit{pawprint}. In order to fill the gap between the detectors and get a continuous area of 1.5 deg$^2$, called a \textit{tile}, six exposures are required.

The VHS survey \citep{McMahon13} is one of the six first generation public surveys carried out on VISTA and is imaging the southern hemisphere (i.e. about 18,000 deg$^2$) with the exception of the area already covered by the VISTA-VIKING and VISTA-VVV surveys. At least two wavebands ($J$ and $Ks$) are being imaged with a 5$\sigma$ limiting magnitudes of $J_{AB}=21.1$ mag and $Ks_{AB}=19.9$ mag respectively, which is $\sim$30 times fainter than the Two Micron All Sky Survey (2MASS). A smaller area of $\sim 5000 \mathrm{deg^2}$ in the South Galactic Cap, which overlaps with the Dark Energy Survey \citep[DES,][]{Abbott18}, is observed more deeply and includes also some coverage in the $H$-band (for observations up to 2012). The VHS survey can be divided into three regions:

\begin{enumerate}
	\item VHS-DES ($\sim 5000 \mathrm{deg^2}$) with deeper observations (median $5\sigma$ $J$-band limiting depth = 21.2 mag);
	\item VHS-GPS which is covering around the Galactic Plane (Galactic latitude $-30 < b < +30\degr$; median $5\sigma$ $J$-band limiting depth = 20.5 mag);
	\item VHS-ATLAS for the remaining area (at $b > +30\degr$ and $b < -30\degr$; median $5\sigma$ $J$-band limiting depth = 20.7 mag).
\end{enumerate}

One of the main scientific aims of the VHS survey is to search for high-redshift quasars, due to its deep and wide coverage.\\
The work presented here uses an internal release of VHS data, processed by the Cambridge Astronomy Survey Unit \citep[CASU;][]{Emerson04} and covering observations taken from 2009 November to 2017 March (i.e ESO Period 84 to 98). We used the aperture corrected magnitude in a 2" diameter (\texttt{apermag3}) as it is the most reliable magnitude for point-source objects, giving the best signal-to-noise in typical seeing conditions.\footnote{\url{http://horus.roe.ac.uk/vsa/dboverview.html}}

\subsection{The Pan-STARRS1 3$\pi$ survey}

The Pan-STARRS1 (PS1) 3$\pi$ survey used a 1.8m diameter telescope located at the Haleakala Observatories in Hawaii. It covers, over four years, the whole sky up to declination $>-30\deg$, i.e. more than 30,000 $\mathrm{deg^2}$, in five imaging filters \citep[$g_{P1},\ r_{P1},\ i_{P1},\ z_{P1},\ y_{P1}$;][]{Chambers16}. The 5$\sigma$ point-source limiting magnitudes are 23.3, 23.2, 23.1, 22.3, 21.4 mag respectively. 
The data used here are from the first public catalogue (DR1, StackObjectThin table), released in 2016 December and which contains about 3.5 billions rows. In all this paper the Pan-STARRS magnitudes cited referred to the PSF magnitudes (\texttt{psfmag}).

\subsection{The \textit{Wide-field Infrared Survey Explorer}}
\label{sec:WISE}

The WISE survey \citep{Wright10} has observed the entire sky in four wavebands W1, W2, W3 and W4 with central wavelengths of 3.4, 4.6, 12 and 22$\micron$ and 5$\sigma$ limiting AB magnitudes of 19.2, 18.8, 16,3 and 14.5 mag respectively. An all-sky catalogue (WISE All-Sky) containing over 563 million objects was released in 2011 March and includes data taken between 2010 January to August. 

A post-cryogenic phase \citep[NEOWISE;][]{Mainzer11} then conducted observations of 70\% of the sky at 3.4 and 4.6$\micron$ between 2010 September and 2011 February. Both the All-Sky and NEOWISE surveys were combined to create the AllWISE Source Catalogue which contain over 747 millions objects and was released in 2013 November. A third phase (NEOWISE Reactivation) observing also at 3.4 and 4.6$\micron$ is still ongoing since 2013 December.

The angular resolution (FWHM) is 6.1$\arcsec$, 6.4$\arcsec$, 6.5$\arcsec$ and 12.0$\arcsec$ for bands W1 to W4. The AllWISE coadds have a size of 4095 x 4095 pixels with a pixel scale of 1.375$\arcsec$.

A new set of unblurred WISE coadds, called unWISE \citep{Lang14} are also available and have a size of 2048 x 2048 pixels with 2.75$\arcsec$/pixel. They have a better signal-to-noise ratio (S/N) than the Atlas images from the AllWISE data products. While the Atlas images are optimised for isolated point source detection, the unWISE coadds, which have a better resolution, are more suitable for forced photometry as we already know the position of the sources. We used the unWISE coadds in this work (see Section \ref{sec:WISE_fp}).

\section{Quasar candidates selection}
\label{sec:QSOsel}

The main spectroscopic feature of a high-z ($z>6$) quasar is the spectral depression or continnum break shortward of the hydrogen Ly-$\alpha$ line ($\lambda_{rest}=1215.67\mathrm{\AA}$), which is redshifted into the optical-NIR range with $\lambda_{obs}\geqslant8500\mathrm{\AA}$. At $z\sim6.5$, the Ly-$\alpha$ line is observed around $9100\mathrm{\AA}$, corresponding to the $z$-band, which leads to red $(z-y)$ and $(z-J)$ colours. At these redshifts, the optical and near IR colours of quasars are similar to those of late-type stars (M, L, T spectral types), which represent the main contaminants in our search. High-z ($z>6$) quasars are rare compared to M-stars and LT type brown dwarfs, making them very difficult to find with reliability.

Our selection of quasar candidates is based on the combination of optical and NIR colours to identify potential high-z quasars followed by candidate classification. The classification is based on spectral energy distribution (SED) fitting that uses optical, near IR and mid IR photometry in order to reliably distinguish between high-z quasars and the more numerous M-stars and brown dwarfs.

\subsection{Catalogue cross-matching and colour selection}
\label{sec:colourSelection}

\begin{figure}
    \includegraphics[width=\columnwidth, trim={0cm, 0cm, 1cm, 1.2cm}, clip]{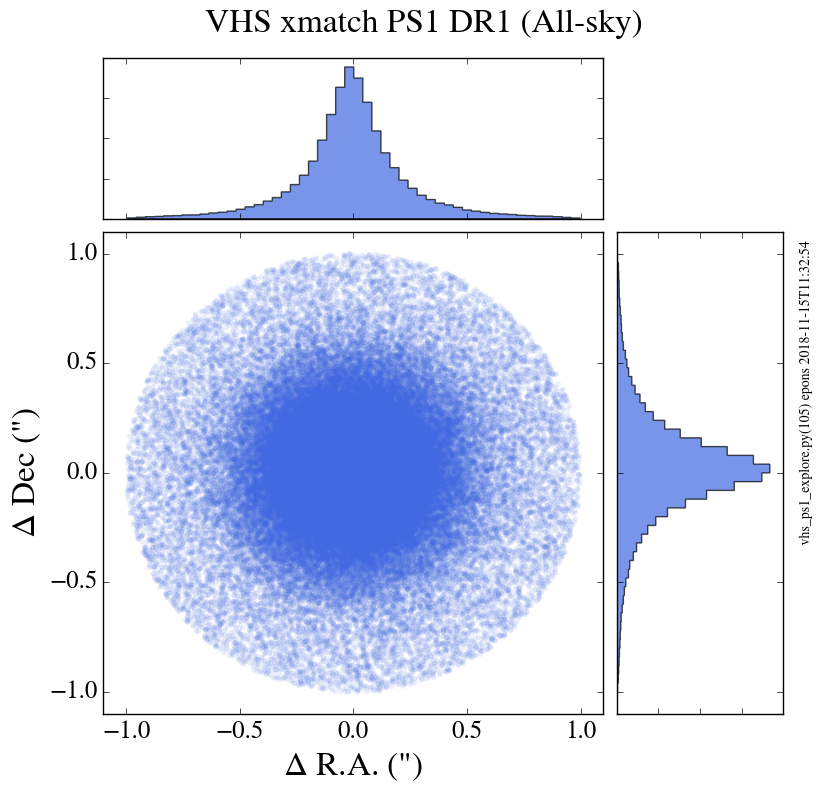}
    \caption{Positional offsets between the VHS and Pan-STARRS position using a match radius of $1\arcsec$ for our all-sky matched catalogues. The offsets in R.A. and Dec are $-0.009\arcsec$ and $0.013\arcsec$ respectively; with a scatter of $\sigma=0.17\arcsec$ for both.}
    \label{fig:offsets}
\end{figure}

We start with the VHS catalogue and look for the closest counterparts in PS1 with a search radius of $1\arcsec$. We get a good agreement between the positions in VHS and PS1, with a median offset of $-0.009\arcsec$ in R.A. and $0.013\arcsec$ in Dec; with a scatter of $\sigma=0.17\arcsec$ for both R.A. and Dec (see Figure \ref{fig:offsets}). 

In both VHS and PS1 we only consider primary detection objects (i.e. flagged non-duplicate detections of the same source) which in principle strongly reduces the number of duplicates. In the VHS-GPS area we exclude objects near the Galactic plane with $-8 < b < +8\degr$ (for $0 < R.A. < 150\degr$) and for the area corresponding to the Galactic bulge $-10 < b < +10\degr$ ($250 < R.A. < 360\degr$). We then cross match with WISE counterparts from the AllWISE catalogue within $3\arcsec$ of the VHS position.

Then to get $z$-dropout objects we select sources with red $z-J_{AB}$ colours or those that are undetected in the $z$-band but still detected in the $J-$band:

\begin{equation}
    \begin{split}
        & z-J_{AB} \ga 1.5\: \text{mag} \: \text{OR}\\
        & z\: \text{NULL AND}\: J_{AB}\: \text{not NULL} 
    \end{split}
\end{equation}
We use a $z-J_{AB}$ limit of 1.5 magnitudes to get drop out factor of 4 in the z waveband compared to J (i.e. $2.5\log(4.0)$). A comparison of the the $z-J$ colours of know quasars indicates that $z-J>1.5$ recover 4 out of the 5 known $z>6.5$ quasars with Pan-STARRS photometry. This still leaves us with a very large number of objects (over 4 million sources ; $\sim$1\% of the input catalogue) for the whole Southern Hemisphere with Dec $ < -30\degr$.

To restrict the number of candidates, we remove very bright objects in the $J$-band which are unlikely to be quasars and those with a low S/N (i.e. a large error in the $J$-band):

\begin{equation}
     J_{AB} < 16 \: \text{mag} \: \text{AND}\: \sigma_{J} \ga 0.15 \:\text{mag}
\end{equation}

The chosen threshold of 16 mag in the J-band is consistent with the observation that the brightest known quasar at $z>6$ has $J_{AB} \sim 18$ mag \citep[SDSS J0100+2802 at $z=6.3$ discovered by][]{Wu15} and because all of the SDSS DR12 quasars have a UKIDSS $J_{AB} > 16.6$ mag \citep{Paris17}. 1.4 million sources remain after the above step. Since high redshift $z$-dropout selected quasars should be undetected in the bluer wavebands, we also required the object to be undetected in all the three optical $g$, $r$, $i$ bands to get a true $z-$dropout (i.e. null detection or fainter than the Pan-STARRS $5\sigma$ magnitude limit):

\begin{equation}
    \begin{split}
        & g, r, i\: \text{NULL}\: \text{OR}\\
        & g,r,i > 23\: \text{mag} \; \text{AND}\: \sigma_{g, r, i} > 0.2\: \text{mag}
    \end{split}
\end{equation}

The above step reduces the number of candidates by a further factor of
$\sim$10 to $\sim$180,000 candidates.

\begin{table}
	\centering
	\caption{Number of objects left in our VHS-PS1 catalogue after the different selection steps, for the three VHS area (ATLAS, DES and GPS) and in total.}
	\label{tab:qso_sel}
	\begin{tabular}{lcccc} 
		\hline
		 & ATLAS & DES & GPS & Total\\
		\hline
		$J$-band detected & $6.6\times 10^7$ & $2.4\times 10^7$ & $1.7\times 10^8$ & $2.6\times 10^8$ \\
	        \hline
		\multicolumn{5}{|c|}{Colour selection (Section \ref{sec:colourSelection})}\\
		\hline
		$z$-dropouts           & 939,149 & 521,618 & 2,744,652 & 4,205,419\\
		$J$-band                & \multirow{2}{*}{282,197} & \multirow{2}{*}{176,133} & \multirow{2}{*}{916,419}      & \multirow{2}{*}{1,374,749}\\
		$\quad$selection   & & & & \\
		g, r, i                       & \multirow{2}{*}{39,244}  &  \multirow{2}{*}{22,477}  & \multirow{2}{*}{114,390}       & \multirow{2}{*}{176,111}\\
		$\quad$undetected & & & & \\
		PS1 junk             & \multirow{2}{*}{29,115}  & \multirow{2}{*}{17,515}   & \multirow{2}{*}{60,939}         & \multirow{2}{*}{107,569}\\
		$\quad$removal    & & & & \\
		\hline
		\multicolumn{5}{|c|}{SED fitting selection (Section \ref{sec:chi2selection})}\\
		\hline
		photo-$z$ \& $\chi^2$ & 1,259 & 1,207 & 2,224 & 4,690\\
		Visual checks & 269 & 500 & 334 & 1,103\\
		\hline
		\multicolumn{5}{|c|}{NTT selection (Section \ref{sec:spectroObs})}\\
		\hline
		$y < 21$ mag & 115 & 132 & 192 & 439\\
		\hline
	\end{tabular}
\end{table}

\subsection{Pan-STARRS false-null junk removal}

At this stage, after the initial colour selection we end up with $\sim 180,000$ sources but based on visual inspection a lot of them are spurious detections due to limitations in the data processing, artifacts (e.g. optical ghosts, cosmic rays ; remove by visual inspection, see end of Section \ref{sec:chi2selection}) or have unreliable photometry (e.g. null values in the catalogue).

In this section we identify the false-null detection in the Pan-STARRS catalogue, i.e. sources  classified as undetected in $g$, $r$, $i$ (i.e. no measured flux in the catalogue) while indeed there is an object present in the images. To remove these false null sources we computed the median and
the $\sigma_{MAD}$ ($\sigma_{MAD}$ is a robust estimator of
  the Gaussian sigma where $\sigma_{MAD} = 1.4826 \times \text{MAD}$) in a box of 80 pixels, corresponding to 20" ($cutout$) around the source position,
that we compared with the pixel mean values in a central box of 8 pixel
($window$). The noise excess in the small 8 pixels (i.e. $2\arcsec$) window
is computed as:

\begin{equation}
    \text{noise \:excess}_{window} = \frac{\text{mean}_{window} - \text{median}_{cutout}}{\sigma_{MAD, cutout}}
\end{equation}

We have compared the noise excess in a test sample of 1500 objects with the
visual check of the PS1 cutouts to derive an empirical threshold for the
removal of false-null junk. The noise excess threshold we
used to remove junk is 0.46, 0.65, 0.63 for the $g$, $r$ and $i$-bands
respectively. This thresholding reduces the candidate lists by about 40\% to
$\sim$110,000 objects. The number of objects left after
each selection step, for the three VHS regions (i.e. ATLAS, DES and GPS)
are given in table \ref{tab:qso_sel}.

\subsection{WISE forced photometry}
\label{sec:WISE_fp}

In order to better distinguish between brown-dwarfs and high-z quasars in the SED fitting we performed forced photometry on WISE for sources not present in the AllWISE catalogue. We used the unWISE coadds \citep[NEO2 release;][]{Lang14}, which are more appropriate for list driven forced photometry. UnWISE coadds have a pixel scale of 2.75$\arcsec$ compared to 1.375$\arcsec$ for the AllWISE Atlas Images (see Section \ref{sec:WISE}).

To do the forced photometry we used \texttt{IMCORE\_LIST}\footnote{\url{http://casu.ast.cam.ac.uk/surveys-projects/software-release/imcore}} from the \texttt{CASUTOOLS}, with the VHS position as an input to extract the WISE flux. The aperture radius was set to 2 pixels, corresponding to $5.5\arcsec$. Due to the PSF of WISE, a large aperture radius was needed to include most of the flux, even if it may be contaminated by nearby sources. A smaller radius aperture could in principle be used to reduce contamination but this would be more sensitive to pixel level quantisation effects depending on whether the source was centred on the centre or edge of a WISE pixel. Further work is needed to determine the optimal solution. An empirical aperture corrections (of 0.315 mag for W1 and 0.408 mag for W2) was applied to the measured WISE magnitudes computed from the median between the forced photometry magnitudes and the AllWISE profile-fit magnitudes (mpro). By considering only point-source objects with reliable photometry (i.e. magnitude errors $<0.1$ mag), we get a small scatter
  in the offsets between the forced photometry and the AllWISE mpro magnitudes
  of 0.010 and 0.011 mag for W1 and W2 respectively.

\subsection{Spectral Energy Distribution Fitting}

The fitting method we used in this work for VISTA, PS1 \citep{Tonry12} and WISE filters was initially developed by \citet{Reed17}.
Photometric data, covering the optical, NIR and MIR, are fitted with both
quasars and brown dwarfs models. We improved both models, by
incorporating a larger set of quasar models and
using the same photometric filters
as the data, to better distinguish between high-z quasars and brown dwarfs.

\subsubsection{The quasars templates}

We used several quasar templates \citep{Maddox12} with different redshifts,
ranging from 5 to 8 with an redshift increment of 0.1, and 10 different levels
of reddening ($E(B-V)$ from 0 to 0.14 with an increment of 0.02, plus two
negative reddening of $-$0.02 and $-$0.01 to allow fitting of quasars with
bluer continuum).

\begin{figure}
    \includegraphics[width=\columnwidth, trim={0cm, 0cm, 0cm, 1cm}, clip]{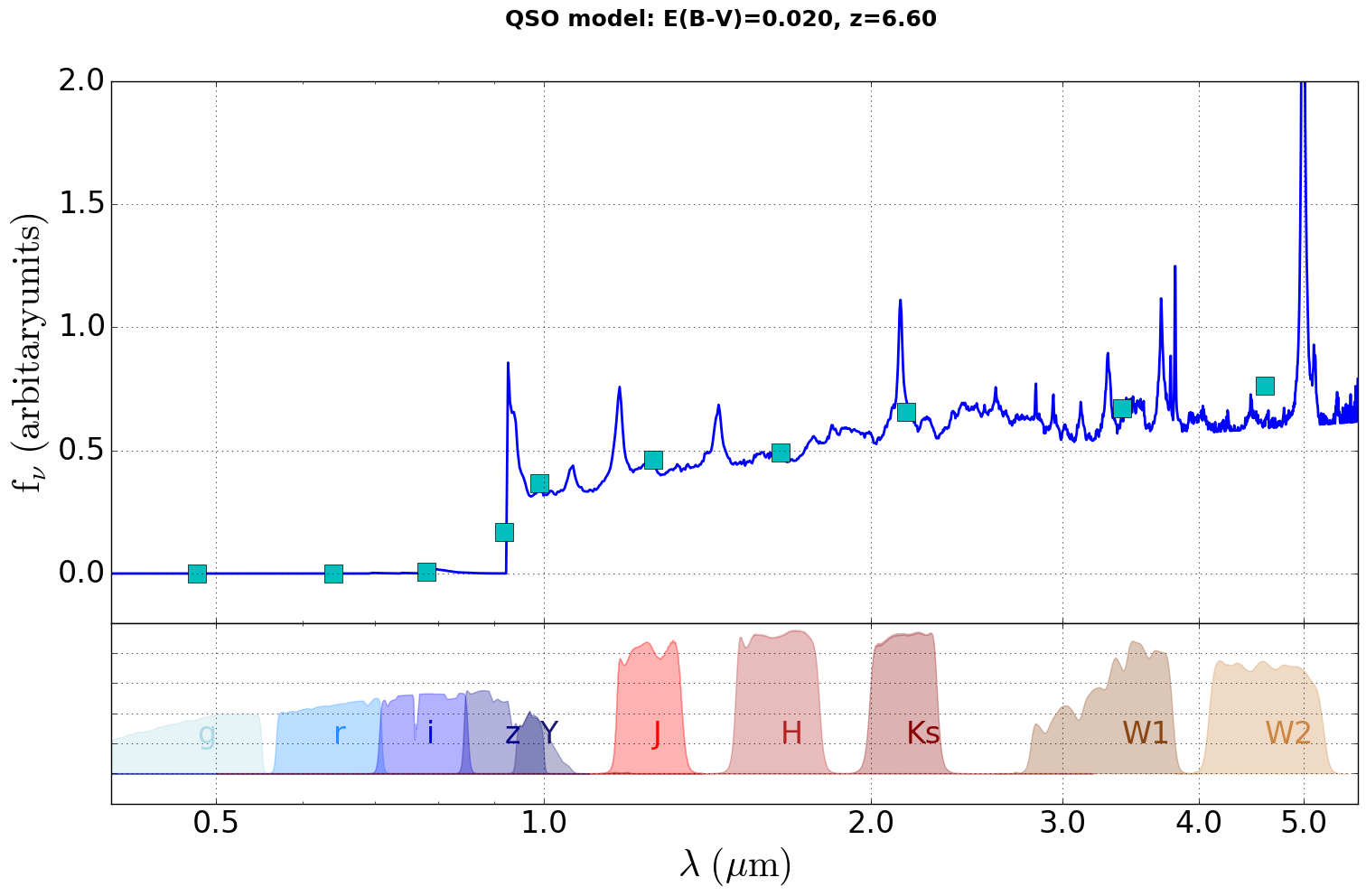}
    \caption{Example of a template of a quasar spectra (blue line) at redshift 6.6 with an intrinsic reddening of $E(B-V)=0.02$. The cyan squares are the integrated flux over each of the PS1, VISTA and WISE filters, which are shown in the bottom panel.}
    \label{fig:QSOtemplate}
\end{figure}

By increasing the range and using a smaller step size in the reddening compared to \citet{Reed17}, we better constrain the best fitting model and get a more accurate fit with a better estimation of the quasar reddening and redshift. In future, we will consider including some variation in the emission line lengths.  We then integrate the flux of the quasar templates in each of the photometric filters we are interested in (all PS1 and VISTA plus W1 and W2 in WISE). An example is shown in Figure \ref{fig:QSOtemplate} for a redshift 6.6 quasar with an intrinsic reddening of $E(B-V)=0.02$. As the flux of the quasar templates are in arbitrary units, we multiplied the model fluxes by a normalization factor obtained from a weighted linear fit between the data and model fluxes.

\begin{figure*}
    \includegraphics[width=2\columnwidth, trim={0cm, 0cm, 21.6cm, 1cm}, clip]{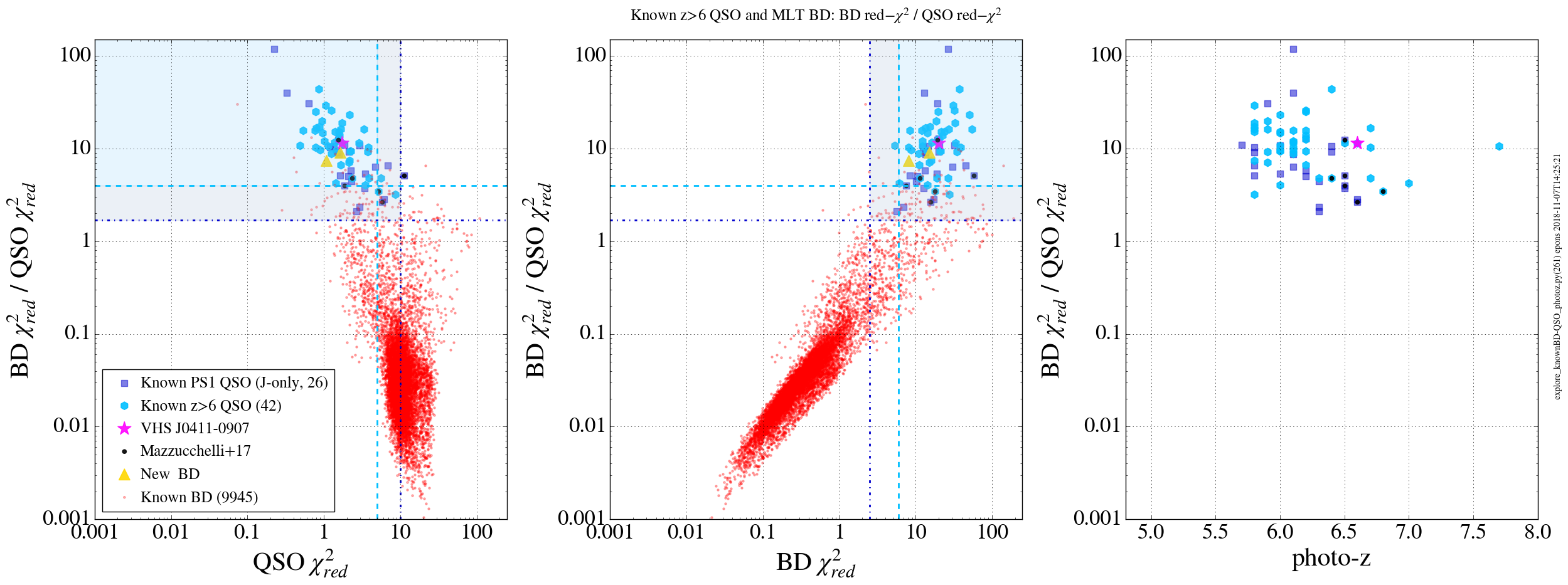}
    \caption{Results of SED fitting reduced-$\chi^2$ for the \citet{Best18} spectroscopically confirmed MLT dwarfs (red dots) and the known $z>6$ DES, PS1 and the two quasars ULAS $z>7$ (blue symbols) The quasars with only the $J$-band available in the NIR are shown by dark blue squares. Our newly discovered quasar VHS J0411-0907 (see Section \ref{sec:vhs0411}) is shown with a magenta star, as well as our new two BD (yellow triangles, see Section \ref{sec:BDmagellan}) and the \citet{Mazzucchelli17} quasars (black points). The blue dashed lines correspond to our quasar candidates selection thresholds (see equations \ref{eq:chi2QSO}, \ref{eq:chi2BD}, \ref{eq:chi2BDQSO}). To distinguish between quasars and cool low mass stars we looked at the $\chi^2_{red,BD} \:/\: \chi^2_{red,QSO}$ ratio vs $\chi^2_{red,QSO}$ (left panel) and vs $\chi^2_{red,BD}$ (right panel).}
    \label{fig:knownQSO-BD}
\end{figure*}

\subsubsection{The MLT dwarfs templates}
\label{sec:BDmo}

For the MLT spectral type dwarf model, we used a new large sample of about 10,000 spectroscopically confirmed MLT brown dwarfs from \citet{Best18}, which have been detected in PS1. The sample of MLT dwarfs are divided in 30 spectral types from M0 to T9. Compared to the SDSS sample from \citet{Skrzypek15} used by \citet{Reed17}, this sample is about 10 times bigger, span a larger range of spectral types and has the advantage of being in the PS1 photometric system. Thus we do not need to compute colour terms to convert from the SDSS photometric system to the PS1 system, which will include some uncertainties, especially as the colour range of normal stars is different from that of brown dwarfs.

In order to get a model for each spectral type, we computed colours in adjacent bands (i.e $r-i$, $i-z$, $z-y$, ..., $W1-W2$) and fit colour vs spectral types with different order (2 to 6) of polynomial functions (see Figure \ref{fig:BDmo} and Table \ref{ap:polyfit} for the coefficients of the polynomial fits). All the magnitudes have been converted to the AB system, and the best fit is selected to have the smallest $\chi^2$ between the data and the fit. We then inferred the magnitude in each filters by assuming a initial magnitude in the $r-band$ of 23. As the \citet{Best18} dwarfs sample has 2MASS photometry in the NIR, we transformed the $J$, $H$, and $Ks$ magnitudes in VISTA magnitudes using the well defined VISTA-2MASS colour equations from CASU \footnote{\url{http://casu.ast.cam.ac.uk/surveys-projects/vista/technical/photometric-properties}}, which have been converted to the AB system.

\begin{figure*}
    \includegraphics[width=2\columnwidth]{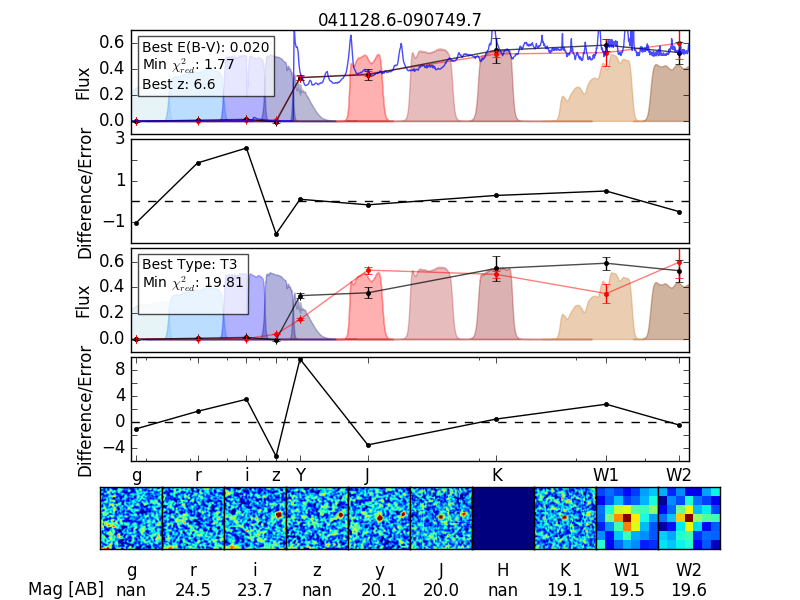}
    \caption{Results of SED fitting for VHS J0411-0907. The top panel shows the best fitting quasar model photometry (in red) to the data (in black). The blue line is for the quasar spectral template used to inferred model photometry. The filled area in the background represent the Pan-STARRS, VISTA and WISE filters. The second panel below, shows the residuals between the data and the fit, divided by the total errors. The two other panel below show the same thing but for the brown dwarf model. Finally in the bottom panel are shown the $20\arcsec$ cutouts in each filters along with the AB magnitudes.}
    \label{fig:SED_VHS0411}
\end{figure*}

\subsubsection{$\chi^2$ selection of high-z quasar candidates}
\label{sec:chi2selection}

We fit the photometric data for all the colour-selected candidates from section \ref{sec:colourSelection} (see also Table \ref{tab:qso_sel}) with the different quasar and brown dwarf templates . The best fit models are those with the smallest reduced-$\chi^2$ ($\chi^2_{red,i}$), which can be computed, for each template $i$ as:

\begin{equation}
    \chi^2_{red,i} = \sum_{n=1}^{N} \left ( \frac{data_n - model_{i, n}}{\sigma(data+model_i)_n} \right )^2 \Bigg/ \;(N-1)
\end{equation}
with N being the number of photometric points and (N-1) the degree of freedom.

Systematic photometric uncertainties on the model have been added in quadrature to the statistical photometric errors, following \citet{Reed17} where it was found that the reduced-$\chi^2$ for the best-fitting quasar and brown dwarf models was often larger than 3, indicating the possibility of systematic errors on measurements or inaccurate SED models. We used percentage errors in flux of 10\%, 10\%, 10\%, 20\%, 5\%, 5\%, 5\%, 5\%, 20\%, 20\% for the $g,\, r,\, i,\, z,\, y,\, J,\, H,\, K,\, W1,\, W2$ bands respectively from \citet{Reed17}.

After selecting the best-fit quasar model and the best-fit brown dwarf model for each candidates, we have selected the most likely high-z quasar candidates based on the photometric redshift and the reduced-$\chi^2$ of the fit. The best discriminant values on the quasar and brown dwarf reduced-$\chi^2$ was empirically derived by fitting the known PS1 and DES high-z quasars (at redshift above 6) from the literature and the \citet{Best18} MLT dwarfs with our SED templates (see Eq \ref{eq:chi2QSO}, \ref{eq:chi2BD}, \ref{eq:chi2BDQSO}). We also included in our analysis the two ULAS quasars at $z>7$ using for the optical bands the DECaLS (forced photometry) magnitudes which uses the DECam camera which is the same as for the DES survey. In order to get more reliable results, we recomputed model fluxes for both the quasar and brown dwarfs templates that correspond to the same filters as the data. 

Our new set of quasar model, with a smaller $E(B-V)$ increments, compared to the quasar model used by \citet{Reed17} with only four level of reddening [$E(B-V)$ = 0.0, 0.025, 0.05, 0.1] gives a better estimation of the classification and a small improvement of the redshift. Indeed, for the known $z>6$ quasars (the same as shown in Figure \ref{fig:knownQSO-BD}), we found a smaller quasar reduced-$\chi^2$ with a median closer to 1 (median $\chi^2_{red, QSO}$ = 1.7 vs 2.1 before), leading to a higher ratio between the BD and quasars reduced-$\chi^2$ (median $\chi^2_{red, BD/QSO}$ = 10.3 vs 7.7 before). Also, we have small improvement on the redshift estimation with a smaller difference between the spectroscopic and photometric redshifts ($\Delta z$ median = 0.08 vs 0.09 with a similar scatter $\sim 0.18$).

In order to select our high-z quasar candidates, we used cuts on the photometric redshift, the best quasar model reduced-$\chi^2$ ($\chi^2_{red, QSO}$; to select possible high-z quasars), the best brown dwarf model reduced-$\chi^2$ ($\chi^2_{red, BD}$; to remove probable brown dwarfs) and on the ratio of the two reduced-$\chi^2$ ($\chi^2_{red, BD/QSO}$; to select objects which are the more likely to be high-z quasars with respect to the brown dwarf model). The photometric redshift can differ up to 0.2 compared to the spectroscopic redshift ($\Delta z$ median = 0.08, $\sigma_{MAD}=0.18$) with quasars at $z >6.5$, so as we are looking for $z >6.5$ quasars we first select object with a photometric redshift $z_{phot} > 6.3$. We choose to use the following thresholds for the reduced-$\chi^2$ (see also Figure \ref{fig:knownQSO-BD}), which reduced the brown dwarfs contamination with minimal exclusion of known quasars:

\begin{equation}
     z_{phot}\: \geqslant \:6.3
\end{equation}

\begin{equation}
   \label{eq:chi2QSO}
    \begin{split}
        \chi^2_{red,QSO} & \leqslant 5 \quad \text{OR}\\ 
                                                             & \leqslant 10 \:\text{(if only one NIR band available)}
    \end{split}
\end{equation}

\begin{equation}
    \label{eq:chi2BD}
    \begin{split}
        \chi^2_{red,BD} & \geqslant 6 \quad \text{OR}\\
                                                          & \geqslant 2.5 \:\text{(if only one NIR band available)}
    \end{split}
\end{equation}

\begin{equation}
    \label{eq:chi2BDQSO}
    \begin{split}
        \chi^2_{red,BD}/\chi^2_{red,QSO} & \geqslant 4 \quad \text{OR}\\
                                                                                       & \geqslant 1.7 \:\text{(if only one NIR band available)}
    \end{split}
\end{equation}

We used relaxed thresholds in the case where only one NIR band is available as with less data points the SED fitting is less well constrained. It is for the same reason that we used also forced photometry on the optical bands (for both the known $z>6$ literature quasars and the \citet{Best18} MLT dwarfs) when there was a non-detection in some bands, to improved the fit. After this stage there are $\sim$4700 candidates remaining. 

To clean the sample and remove spurious and junk sources we performed visual inspection on the optical to MIR cutouts for all the $\sim$4700 candidates. Indeed due to the unusual colour of high-z quasars, we select in addition to cool dwarfs, many objects with unreliable photometry such as diffraction spikes from bright nearby sources, other instrumental or data processing artifacts or corrupted images.  We were then able to reduce the candidate list by about 80\% to $\sim$1100.

\subsubsection{Selection for spectroscopic observations}
\label{sec:spectroObs}

To validate the redshift and quasar classification of the candidates, spectroscopic observations were obtained from the European Southern Observatory New Technology Telescope (NTT) optical spectrograph EFOSC2. With a spectral coverage from 300 to about 1000nm, it can detect the Ly-$\alpha$ line for quasars with redshift $z<7$. In order to be observable by the NTT we restricted our sample to bright source in the $y$-band with $y \leqslant 21$ mag which gives $\sim$450 candidates.

The total number of candidates and the number per VHS area is listed in Table
\ref{tab:qso_sel} for all the steps presented in this section and in the above sections (colour selection on Section \ref{sec:colourSelection} and SED fitting selection \ref{sec:spectroObs}).

We focused our initial spectroscopic follow-up for high-$z$ quasar search
on the region of the southern galactic hemisphere $b < 0\degr$; i.e.
300 $<$ R.A. $<$ 100$\degr$ as it was the region of the sky observable when
our observing time was scheduled in 2017 November 13-16 at ESA, La Silla.

\begin{figure*}
    \includegraphics[width=2\columnwidth, trim={0cm, 0cm, 1cm, 0cm}, clip]{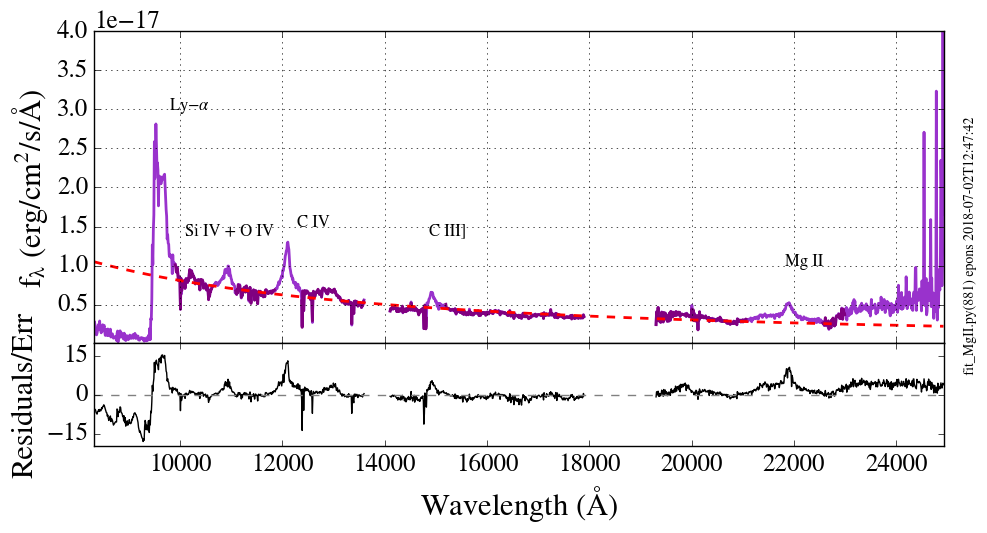}
    \caption{\textit{Top panel:} Magellan FIRE spectrum of VHS J0411-0907. The purple line represents the mean weighted of the data and the black purple line the data used for the continuum fit. The power-law fit of the continuum is shown by the red dashed line. \textit{Bottom panel:} The goodness of the fit is check by looking at the ratio of the difference between the data and the fit (residuals) and the total errors.}
    \label{fig:vhs0411_FIRE}
\end{figure*}

\section{VHS J0411-0907: SED, spectroscopic observations and analysis}
\label{sec:vhs0411}

VHS J0411$-$0907 was the brightest candidate in the VISTA J waveband with
$y<21$ mag. It was considered the most promising high-$z$
quasar candidate for spectroscopic follow-up after visual inspection of
cutouts in all the wavebands (including unWISE NEO2 cutouts)
used for SED fitting.
It has reliable photometry and is significantly detected in PS1 y-band,
VISTA and WISE. See image cutouts in Figure \ref{fig:SED_VHS0411} and
photometry in Table \ref{tab:vhs0411_tab}.

\subsection{Results of the SED fitting}

The photometric redshift was estimated at 6.3 based on the catalogue $riJKsW1W2$ photometry but as there is no flux measured in the $g$ and $z$-band we performed forced photometry on PS1 images to get a better constraint on the redshift, finally predicting it to be 6.6. The SED fit gives a good minimisation of the quasar $\chi^2$ with a $\chi^2_{red,QSO}$ close to 1 and predicts that it is very unlikely to be a brown dwarf due to the high $\chi^2_{red,BD}$ (with value around 20). It is also one of our candidates with the highest $\chi^2_{red,BD}/\chi^2_{red,QSO}$ ratio. The results of the quasar and brown dwarf model fits are shown in Figure \ref{fig:SED_VHS0411} which includes a comparison of the data with the best fit models and 20$\arcsec$ cutouts of the 10 photometric bands. The position and magnitudes of the quasar VHS J0411-0907 are given in Table \ref{tab:vhs0411_tab}.

\begin{table}
	\centering
	\caption{Photometric optical, NIR and MIR magnitudes (in AB) of VHS J0411-0907.}
	\label{tab:vhs0411_tab}
	\begin{threeparttable}
	\begin{tabular}{lc} 
		\hline
		 \multicolumn{2}{c}{Photometric properties}\\
		 \hline
		 R.A. (J2000) & 62.86925 ($04^h11^m28.62\arcsec$)\\
		 Dec (J2000) & $-$9.13048 ($-09\degr07\arcmin49.7\arcsec$) \\
		 $g$    & $>$23.3\tnote{\textdagger} \\
		 $r$     & 24.50 $\pm$ 0.83 \\
		 $i$     & 23.68 $\pm$ 0.29 \\
		 $z$    & $>$22.3\tnote{\textdagger} \\
		 $y$    & 20.09 $\pm$ 0.06 \\
		 $J$    & 20.02 $\pm$ 014 \\
		 $Ks$  & 19.56 $\pm$ 0.21 \\ 
		 $W1$ & 19.48 $\pm$ 0.09 \\
		 $W2$ & 19.59 $\pm$ 0.19 \\
		 \hline
	\end{tabular}
	\begin{tablenotes}
	    \item[\textdagger] Corresponding to the Pan-STARRS $5\sigma$ stack limit magnitude.
	\end{tablenotes}
	\end{threeparttable}
\end{table}

\subsection{Spectroscopic observations}
\label{sec:vhs0411_spec}

The quasar classification and redshift was confirmed in 2017 November using
the optical EFOSC2 \citep{Buzzoni84} spectrograph on the ESO NTT telescope at
La Silla Observatory in Chile. Observations were made with the filter OG530,
the grism \#16 and a slit of $1.5\arcsec$. We took two spectra with a total
exposure time of 3600s.  Data were reduced as in \citet{Reed17} with a custom
set of Python routines and the calibration data were taken during the
afternoon before the observations. From the red edge of the Ly-$\alpha$
emission line we estimated the spectroscopic redshift as 6.81, this is about
0.2 higher than the photometric redshift and it is comparable to the
difference we found for the known quasars at $z>6.5$.

A second spectroscopic observation was conducted in 2017 December
9, (PI: Rob Simcoe), in the NIR, with FIRE, which is a single object, prism cross-dispersed 22 order (orders 11 - 32) infrared spectrometer \citep{Simcoe13} on the Magellan Baade telescope  and which has a spectral resolution of R $=6000$, or approximately $50\;\mathrm{km\;s^{-1}}$, over the range $0.8-2.5 \;\micron$. We observed with a 0.6" slit, and took a spectrum with a 12650s exposure time. The spectrum was reduced using the IDL \texttt{FIREHOSE} pipeline, which performs 2D sky subtraction and extracts an optimally weighted 1D spectrum, as in \citet{Matejek12} and \citet{Chen17}. In order to correct for telluric features and perform flux calibration, spectra of A0V standard stars were obtained as the same time. The calibrated spectrum is fit with a power-law as described in the next section. At wavelengths longward than 23000\AA\ there is an apparent flux excess which may be due to the lower S/N and flux calibration systematic uncertainties in the reddest spectral order of the FIRE  spectrograph.

As we can seen on Figure \ref{fig:vhs0411_FIRE}, in addition to the Ly-$\alpha$, there are four bright emission lines visible: the Si IV + O IV doublet ($\lambda_{rest}=1399.8\mathrm{\AA}$), the C IV line ($\lambda_{rest}=1549.5\mathrm{\AA}$), the C III line ($\lambda_{rest}=1908.7\mathrm{\AA}$) and the Mg II line ($\lambda_{rest}=2799.1\mathrm{\AA}$).

\subsection{Black hole mass and Eddington ratio}

\begin{figure}
    \includegraphics[width=\columnwidth, trim={0cm, 0cm, 0.7cm, 0cm}, clip]{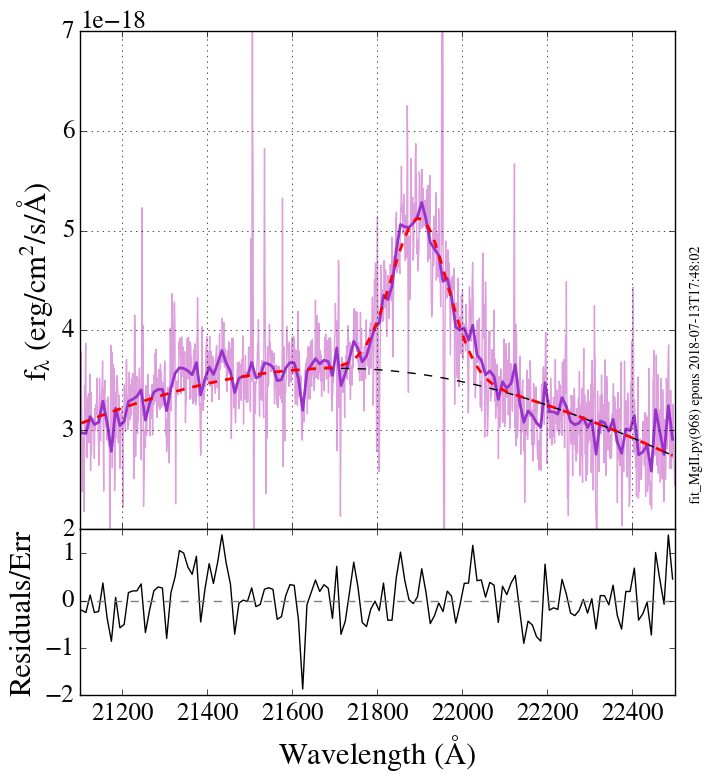}
    \caption{\textit{Top panel:} Magellan FIRE spectrum around the Mg II line (in pink). The total fit of two Gaussians on the mean weighted data (purple line) in shown by the red dashed line. The black dashed line represents the broad Gaussian fit to the strong iron emission-lines around the Mg II line. \textit{Bottom panel:} The two Gaussian fit to reproduce both the Mg II line and the iron lines well reproduced the data as we see small residuals/errors ratio.}
    \label{fig:vhs0411_mgII}
\end{figure}

We use the Mg II $\lambda\lambda$2796, 2804$\mathrm{\AA}$ doublet emission line to estimate the redshift and the black hole mass. The weighted mean of the spectrum was computed using a bin size of $10\mathrm{\AA}$. We first fit the continuum (after masking the emission lines and very noisy data at the red end of the spectrum) with a power-law $f_{\lambda} \propto \lambda^{\alpha}$ (see Figure \ref{fig:vhs0411_FIRE}). We get a spectral slope of $-1.42 \pm 0.02$, which is consistent with the value found for lower redshift quasars \citep{Decarli2010, DeRosa2011} and similar redshift quasars \citep{Mazzucchelli17}. Then, we considered data in a window of $1200 \mathrm{\AA}$ around the Mg II line (see Figure \ref{fig:vhs0411_mgII}), and fit this region with a double Gaussian model. This gives a very good fit with very small residuals.

From the Mg II line fit we found a full-width at half maximum of  $\text{FWHM} = 2112 \pm 94\: \mathrm{km s^{-1}}$ and obtain a redshift of 6.824, slightly larger than from the Ly-$\alpha$ line (corresponding to a velocity offset of $\sim440\: \mathrm{km \: s^{-1}}$). The black hole mass $M_{\rm{BH}}$ is then computed using the relationship from \citet{Vestergaard09}:

\begin{equation}
    \label{eq:MBH}
    M_{\rm{BH}} \:(\mathrm{M_{\odot}}) = 10^{6.86} \left [ \frac{\lambda L_{\lambda}(3000\mathrm{\AA})}{10^{44}\: \mathrm{erg \:s^{-1}}} \right ]^{0.5} \left [ \frac{\mathrm{FWHM_{MgII}}}{10^3\: \mathrm{km \:s^{-1}}} \right ]^2
\end{equation}
with the monochromatic luminosity $\lambda L_{\lambda}(3000\mathrm{\AA})$ obtained from the z=6.8 quasar template normalized using the photometry of our quasar. Using the model instead of the continuum fit to the spectrum reduce the uncertainties related to the calibration, even if in this case we get very similar values ($\lambda L_{\lambda}(3000\mathrm{\AA})=(3.67\pm0.14)\times 10^{46}$ from the quasar template and $\lambda L_{\lambda}(3000\mathrm{\AA})=(3.03\pm0.26)\times 10^{46}$ from the continuum fit).

We also inferred the bolometric luminosity ($L_{bol}$) of the quasar, assuming the bolometric correction of \citet{Shen08}:
\begin{equation}
   \label{eq:Lbol}
    L_{bol} = 5.15\times \lambda L_{\lambda}(3000\mathrm{\AA})
\end{equation}

From the black hole mass ($M_{\rm{BH}}$) we can estimate the Eddington luminosity which is finally used to compute the Eddington ration ($L_{bol} / L_{Edd}$):
\begin{equation}
    L_{Edd} = 1.3\times 10^{38} \left ( \frac{M_{\rm{BH}}}{\mathrm{M_{\odot}}} \right)
\end{equation}

\begin{figure*}
    \includegraphics[width=2\columnwidth, trim={0cm, 0cm, 0.7cm, 0cm}, clip]{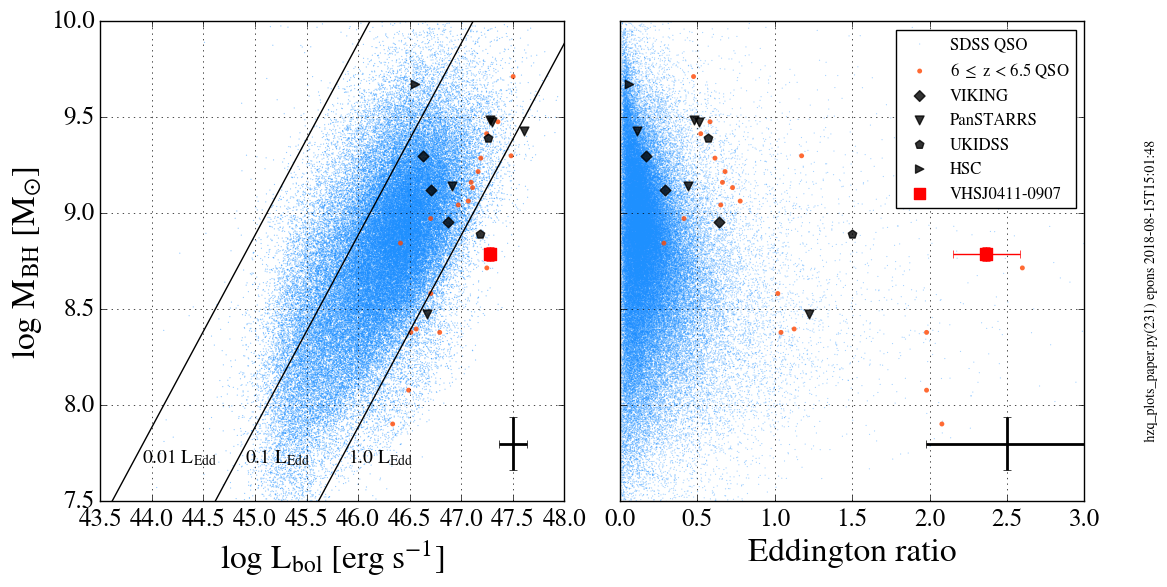}
    \caption{Compaprison of the lack hole masses vs bolometric luminosities (left panel) and Eddington ratios (right panel) for the known quasars at redshift $z>6.5$ (black symbols), the SDSS DR7 quasars (blue points) and the $6<z<6.5$ quasars (orange points). Our new quasar VHS0411-0907 (red square) has one of the lower $M_{\rm{BH}}$ and the highest Eddington ratio compared to other $z>6.5$ quasars. The mean errors for the known quasars at $z>6.5$ are shown in the right bottom corners. On the left panel plot we also show black lines representing constant Eddington luminosity.}
    \label{fig:MBH_knownHzQ}
\end{figure*}

As we can see on Figure \ref{fig:MBH_knownHzQ}, our new quasar VHS J0411-0907 has
  one of the lowest black hole mass and is the highest Eddington ratio quasar discovered at $z>6.5$. We recomputed the black hole masses of the known $z>6.0$ quasars using the \citet{Vestergaard09} correlation for comparison. We also compared with the SDSS DR7 quasars ($z<5$) from the \citet{Shen11} catalogue. The high redshift quasars tend to have a higher bolometric luminosity and Eddington ratio than the lowest redshift quasars but a similar black hole mass. The higher bolometric luminosity (and Eddington ratio) observed for high-$z$ quasars compared to the low-$z$ SDSS quasars is most likely due to selection effect, as we preferentially pick high-luminosity quasars for spectroscopic follow-up. Almost all of the $z>6.5$ known quasars have $J_{AB} < 21.5$ mag (see Figure \ref{fig:J_vs_z}) with $K_{s,AB} \sim J_{AB} - 0.5$ mag, which gives, using equation \ref{eq:Lbol} and the K-band flux as a proxy of $f_{\lambda}(3000\mathrm{\AA})$, a minimum bolometric luminosity of $\log L_{bol} \sim 46.6$ \citep[see also ][]{Mazzucchelli17}. The estimated values of the redshifts, black hole mass, bolometric luminosity and Eddington ratio for VHS0411-0907 are compared with the two bright ($J_{AB} < 20.5$ mag) published quasars at $z > 6.8$ in Table \ref{tab:vhs0411_tab2}.

Using our estimated black-hole mass and Eddington ratio for VHSJ0411-0907 we now try to put 
constraints on the progenitor black-hole seeds that would be plausible for this luminous $z=6.82$ quasar. The time 
taken for a black-hole of mass $M_{\rm{BH}}$ to assemble from a seed of mass $M_{\rm{BH,seed}}$ assuming it is 
accreting at a constant Eddington ratio, can be written as:

\begin{equation}
     t\;[\mathrm{Gyr}] = t_s\times \left (\frac{\epsilon}{1-\epsilon} \right ) \times \frac{L_{Edd}}{L_{bol}} \times \ln \left ( \frac{M_{\rm{BH}}}{M_{\rm{BH,seed}}}\right )
\end{equation}

\noindent where $t_s=0.45$ Gyr is the Salpeter time and $\epsilon \sim0.07$ is the radiative efficiency. 
Assuming constant Eddington ratio of 2.37, a $100\;M_\odot$ seed would take 0.22 Gyr to grow to this black-hole 
whereas a $10^5\;M_\odot$ seed would take 0.12 Gyr. So as the age of the Universe at $z=6.8$ is about 0.78 Gyr, both
 low mass and high mass seeds could form a black hole with the observed mass if growing at a Eddington ratio of 2.37. 
 However, it is unlikely that such high levels of accretion can be sustained for this long period of time, so instead we 
 calculated the fixed Eddington ratio required assuming growth at this rate over the entire age of the Universe; it still 
 required a high Eddington ratio of 0.68 and 0.38 for a black hole mass seed of $100\;\mathrm{M_\odot}$ and $10^5\;
 \mathrm{M_\odot}$ respectively. Therefore the high-level of accretion observed in this $z=6.82$ quasar allows for 
 growth from relatively low-mass seeds if it can be sustained over $\sim 0.22$ Gyr. In contrast, a recent study 
 \citep{Mazzucchelli17} report that very massive seeds ($\sim 10^6\;\mathrm{M_\odot}$) are always required to form the 
 observed SMBH in other $z>6.5$ quasars based on an average Eddington ratio of 0.39. Our results can still be 
 consistent with high-mass BH seeds if super Eddington accretion is episodic but we cannot rule out low-mass seed 
 model. Two of the highest-$z$ quasars with high luminosity (our new quasar VHS J0411-0907 and ULAS 
 J1342+0928) have very high Eddington ratio (2.4 and 1.5 respectively) consistent with super-Eddington accretion 
 phases.

\begin{table*}
	\centering
	\caption{Quantities derived from the fit on the NTT and Magellan spectra for VHS J0411-0907 and comparison with values for published bright quasars at $z>6.8$.}
	\label{tab:vhs0411_tab2}
	\begin{threeparttable}
	\begin{tabular}{ccccccccc} 
		 \hline 
		 & \multirow{2}{*}{$z_{Ly-\alpha}$} & \multirow{2}{*}{$z_{MgII}$} & $\lambda L_{3000\AA}$ & FWHM$_{MgII}$ & $L_{bol}$  & $M_{\rm{BH}}$ & \multirow{2}{*}{$L_{bol}/L_{Edd}$} & References \tnote{a}\\
		 & &  &  $(10^{46}\:erg \:s^{-1})$ & ($\mathrm{km\: s^{-1}}$) & $\mathrm{(10^{47}\:erg \:s^{-1})}$ & $(10^8\:M_{\odot})$  & &  disc./fit\\
		 \hline \hline
		 VHS J0411-0907 & 6.81   & 6.824  & $3.67 \pm 0.14$ & $2103 \pm 85$ & $1.89 \pm 0.07$ & $6.13 \pm 0.51$  & $2.37 \pm 0.22$ & [1]/[1]\\
		 \hline
		 ULAS J1120+0641& - & 7.087 & $3.6^{+0.4}_{-1.4}$ & $4258 ^{+524} _{-395}$& $1.83^{+0.19}_{-0.072}
$ & $24.7^{+6.2}_{-6.7}$ & $0.57^{+0.16}_{-027}$ & [2]/[4] \\[5pt]
		 ULAS J1342+0928 & - & 7.527 & 2.97 & $2500^{480}_{-320}$& 1.53 & $7.8^{+3.3}_{-1.9}$ & $1.5^{+0/5}_{-0.4}$ & [3]/[3]\\[5pt]
		 \hline
	\end{tabular}
	\begin{tablenotes}
	    \item[a] References of the discovery and for the parameters derived from spectral fitting (i.e. the $\lambda L_{3000\AA}$, FWHM$_{\mathrm{MgII}}$ and $L_{bol}$): [1] this work, [2] \citet{Mortlock11}, [3] \citet{Banados18}, [4] \citet{Mazzucchelli17}.
	\end{tablenotes}
	\end{threeparttable}
\end{table*}

\section{Current limitation of the method: two new brown dwarfs}
\label{sec:BDmagellan}

A further two candidates were selected for spectroscopic observations with FIRE on Magellan due to slightly higher photometric redshift and $y>21$ and were considered to faint and beyond the redshift senstivity of the NTT with EFOSC since Lyman-$\alpha$ has an observed wavelength $>$9700\AA. They were observed on the same date as VHS J0411-0907 (2017 December 9, PI: Rob Simcoe), with an exposure time of 3815s and 2409s for VHS J0303$-$1941 ($03^h03^m48.73\arcsec$ $-19\degr41\arcmin19.1\arcsec$) and VHS J0325-1116 ($03^h25^m250.93\arcsec$ $-11\degr16\arcmin23.4\arcsec$) respectively; and they spectra were reduced as described in Section \ref{sec:vhs0411_spec}.

These two SED selected quasar candidates with
photometric redshifts close to 7 and high $\chi^2_{red,BD} \:/\: \chi^2_{red,QSO}$ ratios (larger than 8) were spectroscopically confirmed to be brown dwarfs. The Magellan FIRE spectra of these two objects, VHS J0303$-$1941 and VHS J0325-1116, are shown in Figure \ref{fig:BD_FIRE} and the SED fits are shown in Appendix \ref{ap:BD_SED}. The spectra correspond to a early T-type dwarf (T0) and an early L-type dwarf (L2) respectively but our SED brown dwarf model is not able to give a good fit from the photometric data. It may be due to the intrinsic dispersion of the colours of brown dwarfs (see Figure \ref{fig:BDmo}) due to photometric errors and intrinsic spectral diversity. For example, there is a class of T dwarf stars with J-band flux reversal \citep{Looper08}.

A more sophisticated approach to the systematic uncertainties in the models is needed than the approach described in section \ref{sec:chi2selection} (e.g. each spectral type needs a different series of photometric uncertainties). So there are still some limitation on our SED fitting models and we are working on improving the models to get a cleaner list of quasar candidates.

\begin{figure}
    \includegraphics[width=\columnwidth, trim={0cm, 0cm, 0.7cm, 0cm}, clip]{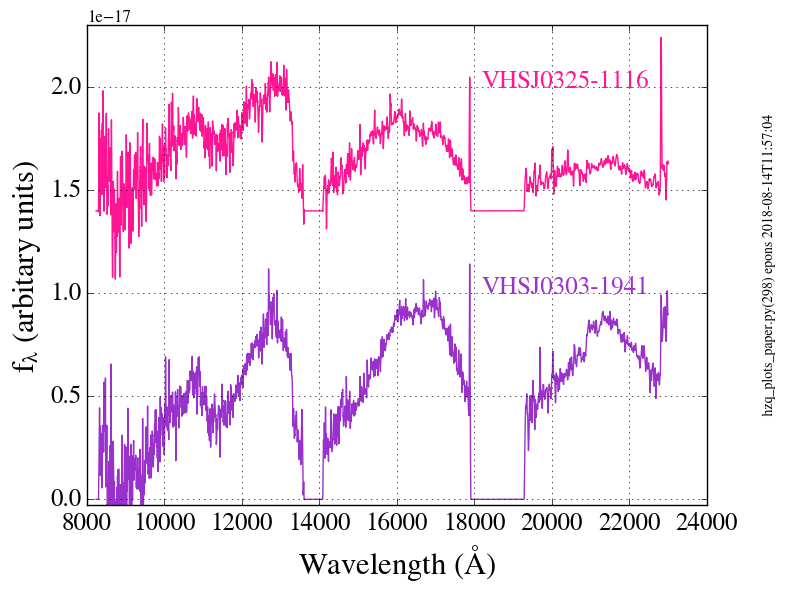}
    \caption{Magellan FIRE smoothed spectra of two brown dwarfs for which the SED fitting was giving a high-redshift quasar classification.}
    \label{fig:BD_FIRE}
\end{figure}

\section{Predictions on the number of high-$z$ quasars}
\label{sec:numQSO}

Currently 6 quasars at redshift $z \ga 6.5$ have been detected in VHS:
our new VHS J0411-0901, four from \citet[in preparation]{Reed17} found in
DES (but not in the PanSTARRS area) plus one from \citet{Mazzucchelli17}
discovered in Pan-STARRS (see Figure \ref{fig:J_vs_z}). Our new quasar and
the Pan-STARRS quasars are in the VHS-ATLAS region (at $b < -30\degr$ and
$b > +30\degr$ respectively) and the DES quasars are in the VHS-DES area.

\begin{figure}
    \includegraphics[width=\columnwidth, trim={0cm, 0cm, 0.8cm, 0cm}, clip]{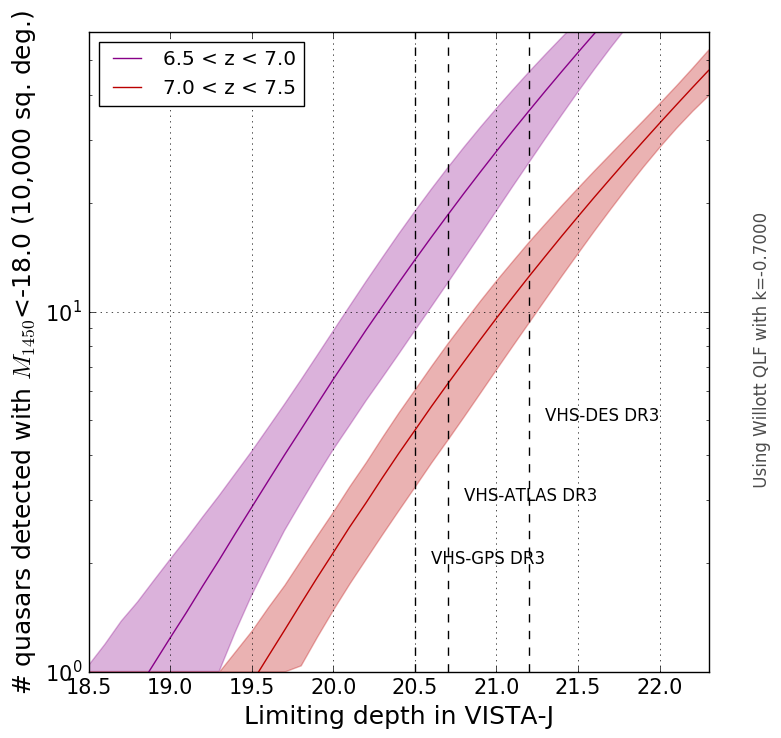}
    \caption{Predicted number of quasars at $z>6.5$ in VHS for 10,000 deg$^2$ based on the $J$-band limiting depth for the three VHS area (GPS, ATLAS and DES, vertical dashed line left to right). The predicted numbers are based on the \citet{Jiang16} luminosity function and are given in two redshift range: $6.5<z<7.0$ (purple curve) and $7.0<z<7.5$ (red curve). The filled area around the curve correspond to $1\sigma$ errors.}
    \label{fig:NumQSO}
\end{figure}

Based on the \citet{Jiang16} luminosity function, we expect to detect in 10,000 deg$^2$ of VHS about 20, 34 and 15 quasars with $6.5<z<7.0$ for the $J$-band limiting depth of VHS-ATLAS, VHS-DES and VHS-GPS respectively. In addition to about 6, 14 and 5 quasars with $7<z<7.5$ (see Figure \ref{fig:NumQSO}). The corresponding numbers for the full area observed by VHS and for the VHS area covered by Pan-STARRS are given in Table \ref{tab:NumQSO}.

\begin{table}
	\centering
	\caption{Number of predicted quasars at $z>6.5$ in VHS and in VHS covered by Pan-STARRS (VHS catalogue for observations up to 2017 March).}
	\label{tab:NumQSO}
	\begin{tabular}{cccc} 
		 \hline 
		                    & VHS-ATLAS & VHS-DES & VHS-GPS \\
		  \hline\hline
		 \multicolumn{4}{c}{VHS area (16,560$\mathrm{deg^2}$)}\\
		 \hline
		 $6.5<z<7$ & 10 & 14 & 11\\
		 $7<z<7.5$ & 3 & 6 & 4 \\
		 \hline\hline
		 \multicolumn{4}{c}{VHS area covered by PS1 (8,650$\mathrm{deg^2}$)}\\
		 \hline
		 $6.5<z<7$ & 8 & 5 & 5 \\
		 $7<z<7.5$ & 2 & 2 & 2\\
		\hline
	\end{tabular}
\end{table}

\section{Summary and Conclusions}

\begin{figure}
    \includegraphics[width=\columnwidth, trim={0cm, 0cm, 0cm, 0cm}, clip]{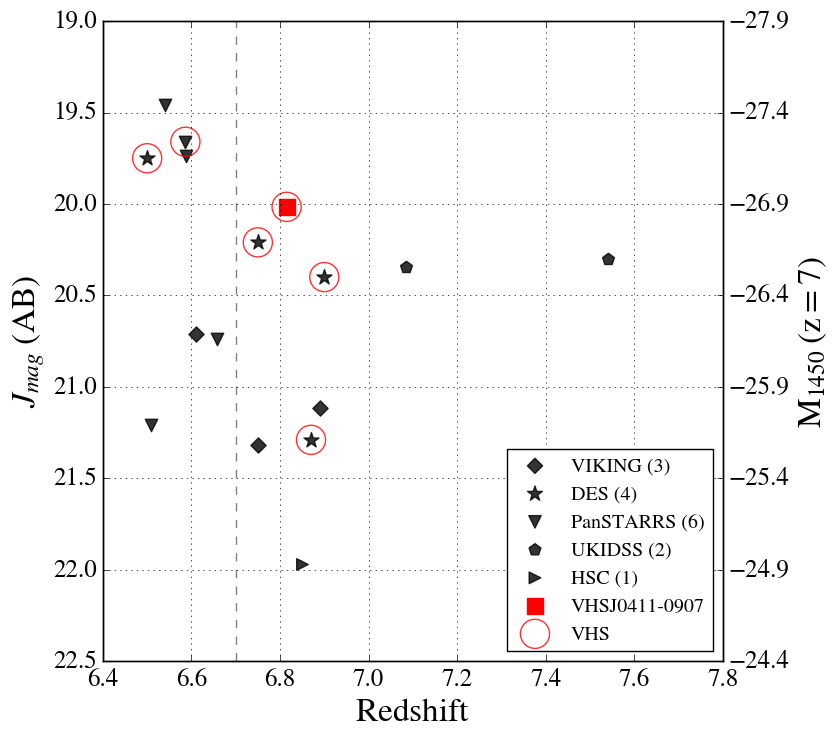}
    \caption{$J$-band magnitude vs redshift for the known quasars at redshift above 6.5 showed in black (Pan-STARRS: down triangles, DES: stars, VIKING: diamond, UKIDSS: pentagon and HSC: right triangle) and for our new quasar VHS J0411-0907 (red square). The objects detected in VHS are circled in blue.}
    \label{fig:J_vs_z}
\end{figure}

By combining colour-selection and SED fitting $\chi^2$ selection we found a new high-$z$ quasar VHS J0411-0907 at a redshift of 6.82. This quasar is the most distant Pan-STARRS quasars and the second highest VISTA bright ($J_{AB}<20.5$ mag); see \citet{Reed19} for information about VDES with $z=6.9$) quasars found at this time and is also very bright compared to other known quasar at similar redshift (see Figure \ref{fig:J_vs_z}). It is indeed the brightest $J-$band quasar at redshift above 6.7.

For the SED based selection we fitted several templates of quasars with different redshifts and reddening and several brown dwarf templates with different spectral types (M, L and T type) and estimated a photometric redshift. We then used the resulting $\chi^2$ of the best fitting quasar and brown dwarf models to classify our colour pre-selected candidates. We used the known quasars at redshift above 6 and the \citet{Best18} spectroscopically confirmed MLT dwarfs to defined the $\chi^2$ selection and find probable high-$z$ quasars candidates. With our $\chi^2$ method we selected five of the six new
  \citet{Mazzucchelli17} quasars, we miss only one which is just above our
  limit in term of $\chi^2_{red,QSO}$ (with $\chi^2_{red,QSO}\sim 11$ and only
  one NIR band).

One of our candidate, VHS J0411-0907, has then been spectroscopically confirmed firstly with the NTT EFOSC2 optical spectrograph and then re-observed with the Magellan FIRE NIR spectrograph. From the fitting of the NIR spectrum we obtained a Mg II redshift and we also inferred the black hole mass, bolometric luminosity and Eddington ratio. Our new quasar VHS J0411-0907 has one of the lowest black hole mass and is the highest Eddington ratio quasar discovered at $z>6.5$. The high Eddington ratio of this quasar is consistent with a scenario of low-mass BH seeds growing at super-Eddington rates.

\section*{Acknowledgements}

We would like to thank the reviewer for its helpful comments.

EP, RGM, MB, SLR and PCH acknowledge the support of UK
Science and Technology research Council (STFC).

Support by ERC Advanced Grant 320596 “The Emergence of Structure during the Epoch of reionization” is gratefully acknowledged.

The analysis presented here is based on observations obtained as part of
the VISTA Hemisphere Survey, ESO Programme, 179.A-2010 (PI: McMahon) and
NTT observations obtained as part of ESO Program ID: 100.A-0346 (PI: Reed).

This publication makes use of data products from the Wide-field Infrared Survey
Explorer, which is a joint project of the University of California, Los
Angeles, and the Jet Propulsion Laboratory/California Institute of Technology,
funded by the National Aeronautics and Space Administration.

This paper makes uses of data products from the Pan-STARRS1 Surveys (PS1) and
the PS1 public science archive, which have been made
possible through contributions by the Institute for Astronomy, the University
of Hawaii, the Pan-STARRS Project Office, the Max-Planck Society and its
participating institutes, the Max Planck Institute for Astronomy, Heidelberg
and the Max Planck Institute for Extraterrestrial Physics, Garching, The Johns
Hopkins University, Durham University, the University of Edinburgh, the Queen's
University Belfast, the Harvard-Smithsonian Center for Astrophysics, the Las
Cumbres Observatory Global Telescope Network Incorporated, the National Central
University of Taiwan, the Space Telescope Science Institute, the National
Aeronautics and Space Administration under Grant No. NNX08AR22G issued through
the Planetary Science Division of the NASA Science Mission Directorate, the
National Science Foundation Grant No. AST-1238877, the University of Maryland,
Eotvos Lorand University (ELTE), the Los Alamos National Laboratory, and the
Gordon and Betty Moore Foundation.




\bibliographystyle{mnras}
\bibliography{refs} 




\appendix

\section{Polynomial fit for brown dwarfs}
\label{ap:polyfit}

\begin{figure*}
    \includegraphics[width=2\columnwidth]{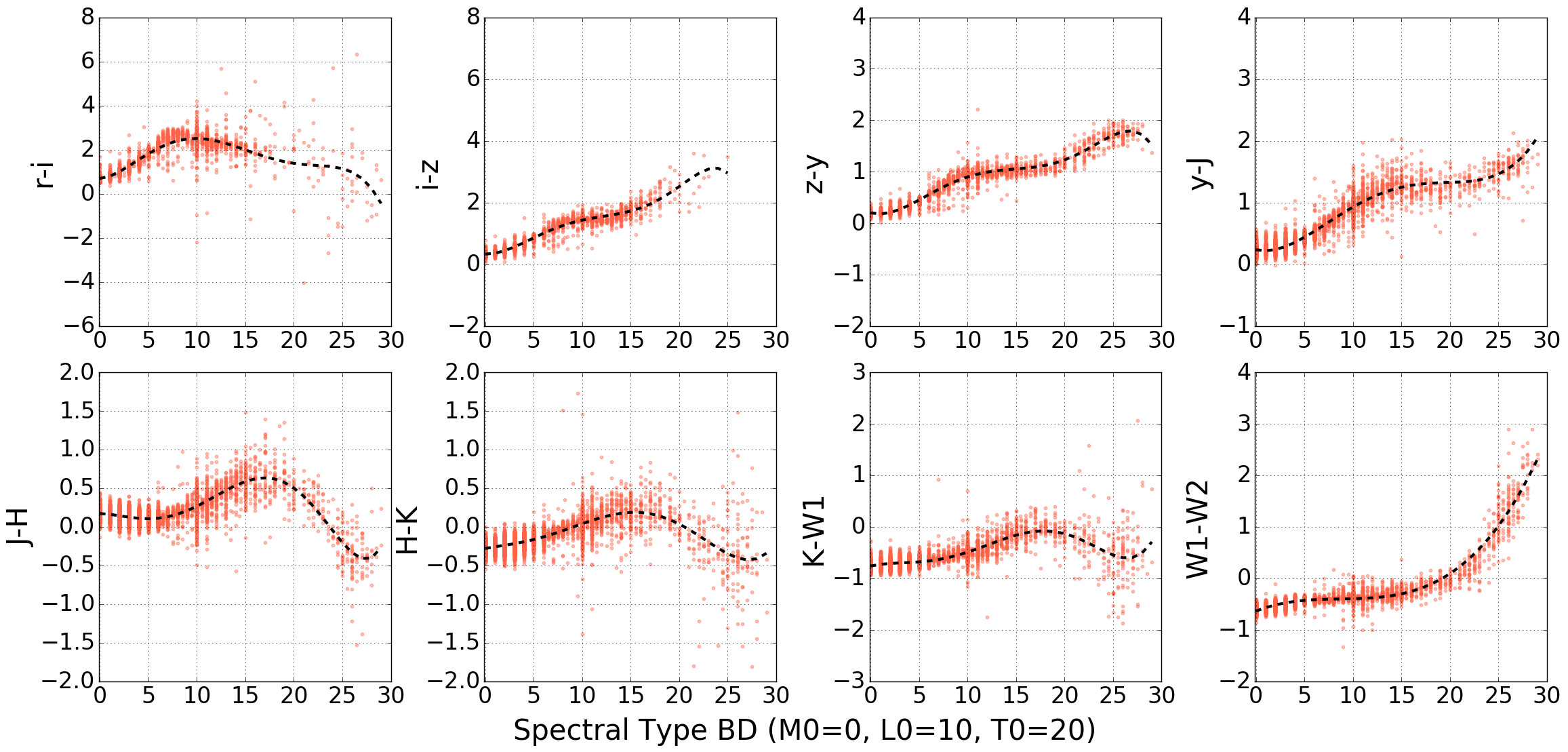}
    \caption{Colour vs spectral type for MLT dwarfs from \citet{Best18} (red points).  The dot black line represents the best polynomial fit to the data.}
    \label{fig:BDmo}
\end{figure*}

We used a sample of spectroscopically confirmed brown dwarfs \citep{Best18} to create our brown dwarf templates needed for the SED fitting. To do this we fit different polynomials of order 2 to 6 on the colours (see section \ref{sec:BDmo}). The coefficients of the best polynomial on each colours (optical and NIR data from Pan-STARRS and 2MASS respectively) is given in Table \ref{tab:pcoeffs} and are shown on Figure \ref{fig:BDmo}.

\begin{table*}
	\centering
	\caption{Coefficients of the best polynomial fit ($p(x) = p6 * x^6 + p5 * x^5 + p4 * x^4 + p3 * x^3 + p2 * x^2 + p1 * x + p0$ with x being the spectral type) on the MLT dwarfs colours.}
	\label{tab:pcoeffs}
        \begin{tabular}{ccccccccc}
coeffs & r-i & i-z & z-y & y-J & J-H & H-K & K-W1 & W1-W2 \\
\hline
p6 &   -               & -2.11e-07   & -6.94e-08  & -                & 5.51e-08    & -                & 5.23e-08   & - \\
p5 &   -6.73e-06 & 8.49e-06    & 2.99e-06   & -                & -1.93e-06   & 1.06e-06   & -1.89e-06  & - \\
p4 &   4.62e-4    & -8.35e-06   & 1.65e-05   & 2.51e-05   & -2.55e-05   & -5.91e-05  & -2.19e-05  & - \\
p3 &   -1.07e-2   & -3.12e-3     & -2.24e-3    & -1.35e-3    & 1.25e-3      & 9.15e-4     & 1.29e-3     & 3.17e-4 \\
p2 &   8.31e-2    & 3.75e-2      & 3.04e-2     & 2.17e-2     & -7.17e-3     & -3.58e-3     & -1.08e-2   & -8.28e-3 \\
p1 &   2.21e-2    & -7.82e-3     & -4.97e-2    & -3.72e-2    & -4.76e-3     & 2.49e-2     & 4.16e-2     & 7.5e-2 \\
p0 &   0.696       & 0.332         & 0.201         & 0.237        & 0.172          & -0.281       & -0.762       & -0.639 \\
\hline
        \end{tabular}
\end{table*}

\section{SED fitting of brown dwarfs}
\label{ap:BD_SED}

\begin{figure*}
    \centering
    \begin{tabular}{cc}
        \includegraphics[width=1\columnwidth, trim={0.7cm, 0cm, 2cm, 0cm}, clip]{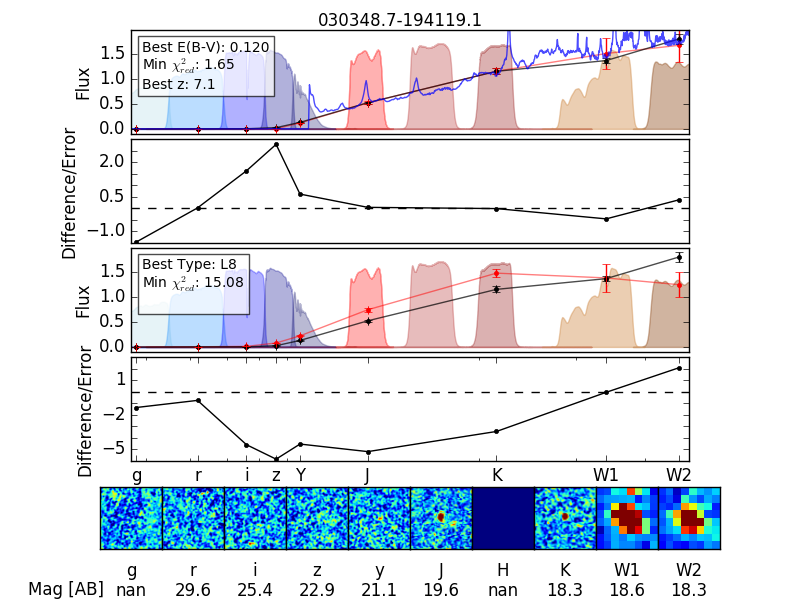}&
        \includegraphics[width=1\columnwidth, trim={0.7cm, 0cm, 2cm, 0cm}, clip]{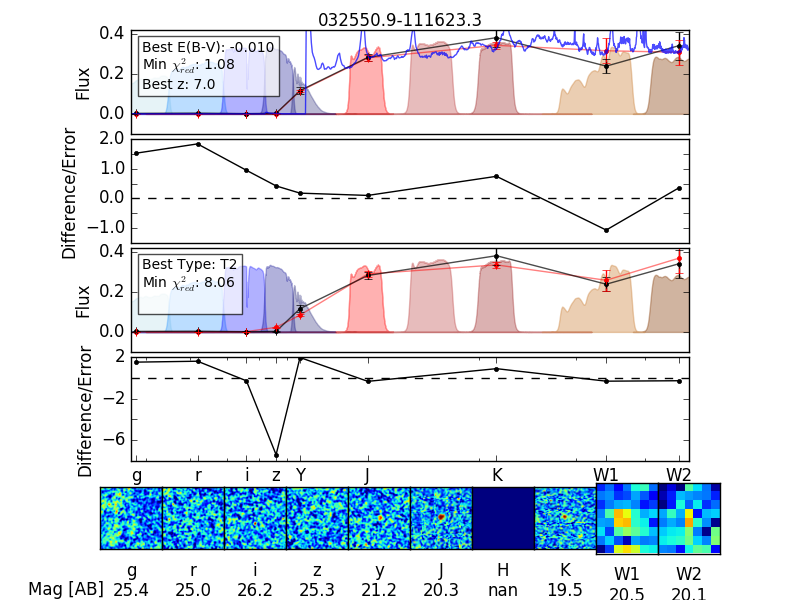}\\
    \end{tabular}
    \caption{SED fitting of two good high-$z$ candidates but spectroscopically confirmed as brown dwarfs. For a full description see Figure \ref{fig:SED_VHS0411}.}
    \label{fig:BD_SED}
\end{figure*}

The SED fitting results of two of our $\chi^2$ selected high-$z$ candidates which was finally spectroscopically confirmed to be brown dwarfs are given in Figure \ref{fig:BD_SED}.

\bsp	
\label{lastpage}
\end{document}